\documentclass[12pt]{article}
\usepackage{cite}
\newcommand{\beq}{\begin{equation}}
\newcommand{\eeq}{\end{equation}}
\newcommand{\Tr}{{\rm Tr}\,}
\newcommand{\nfour}{\mbox{${\cal N}\!\!=\!4\;$}}
\newcommand{\ntwo}{\mbox{${\cal N}\!\!=\!2\;$}}
\newcommand{\none}{\mbox{${\cal N}\!\!=\!1\;$}}

\begin{document}
\begin{titlepage}
\renewcommand{\thefootnote}{\fnsymbol{footnote}}

\begin{flushright}
TPI-MINN-00/25\\
UMN-TH-1905/00\\
PNPI-TH-2396/00\\
ITEP-TH-85/00\\
hep-th/0012250\\
\end{flushright}

\vfil

\begin{center}
\baselineskip20pt
{\bf \Large Type I Superconductivity  upon \\ Monopole Condensation
in Seiberg-Witten Theory}
\end{center}
\vfil

\begin{center}
{\large  A. Vainshtein$^a$ and A. Yung$^{a,b,c}$}

\vspace{0.3cm}

$^a${\it Theoretical Physics Institute, University of Minnesota,
Minneapolis, MN 55455\\[0.2cm]}
$^b${\it Petersburg Nuclear Physics Institute, Gatchina, St. Petersburg,
188350~\footnote{Permanent address}}\\[0.2cm]
$^c${\it Institute of Experimental and Theoretical Physics,
Moscow, 117259}

\vfil

{\large\bf Abstract} \vspace*{.25cm}
\end{center}

We study the confinement scenario in \ntwo supersymmetric SU(2)
gauge theory near the monopole point upon breaking of \ntwo
supersymmetry by the adjoint matter mass term. We confirm claims
made previously that the Abrikosov--Nielsen--Olesen string near
the monopole point fails to be a BPS state once next-to-leading
corrections in the adjoint mass parameter taken into account.
Our results shows that  type I superconductivity arises upon
monopole condensation. This conclusion allows us to make
qualitative predictions on the structure of the hadron mass
spectrum near the monopole point.

\vfil

\begin{flushleft}
December 2000
\end{flushleft}
\end{titlepage}

\section{Introduction}

According to Mandelstam and 't Hooft ideas \cite{1} confinement of
charges appears as a dual Meissner effect upon condensation of
monopoles. Once monopoles condense the electric flux is confined
in the (dual) Abrikosov--Nielsen--Olesen (ANO) vortex \cite{ANO}
connecting the heavy trial charge and anti-charge. The vortex has
a constant energy per unit length (the string tension $T$). This
ensures that the confining potential between the heavy charge and
anti-charge increases linearly with their separation. However,
since dynamics of monopoles is hard to control in
non-supersymmetric gauge theories, 
this picture of confinement remained  an
unjustified qualitative scheme for many years.

The breakthrough in this direction was made by Seiberg and
Witten in \cite{2,3}. Constructing the exact solution of the \ntwo
supersymmetric gauge theory they showed that the condensation of
monopoles really occurs near the monopole point on the
moduli space of the theory once \ntwo supersymmetry is broken
down to \none by the mass term of the adjoint matter \cite{2}.

More specifically, they considered the \ntwo Yang-Mills theory with SU(2) as a
gauge group \cite{2}. The gauge symmetry is broken down to U(1) by the vev
$\langle\varphi\rangle=a\,\tau_3/2$ of the adjoint scalar
field $\varphi$. The complex parameter $a$ is the modulus
on Coulomb branch of the theory.
Near the monopole singularity on the Coulomb branch (the point where
monopoles become massless) the effective low energy theory is
the dual \ntwo QED. This means that the theory has light
monopole hypermultiplet coupled to dual photon multiplet in the
same way as ordinary charges are coupled to the ordinary photon.

The  \ntwo supersymmetry is broken down to \none by adding a
 mass term $\mu\,\Tr\Phi^{2}$ for the adjoint matter
($\Phi$ is the adjoint chiral superfield; its scalar component
$\varphi$ develops the vev discussed above). After the breaking  the whole
Coulomb branch shrinks to two points at which either the monopole or dyon
become massless~\cite{2} (we give a brief review of the phenomenon in
the next section).  Consider, say, the monopole point. At this point  the monopole
condensate develops, its magnitude is proportional to the small parameter $\mu$.
The monopole condensation ensures U(1) confinement of trial heavy charges.

Since the work of Seiberg and Witten \cite{2} it was quite
important to understand to what extent this U(1) confinement of electric
charges  is similar to confinement of color we expect  (but cannot control) in QCD.
Moreover, we expect the QCD-like confinement also in \none
supersymmetric QCD which can be obtained as a large $\mu$
limit of the theory under consideration. Unfortunately, we have
no control on the dynamics of the theory in this limit (with exception of
 values of various chiral condensates which are known exactly at
any $\mu$~\cite{EGR,MK,GVY}).

One distinction noticed by Douglas and Shenker \cite{DS} appears
in SU$(N_c)$ theories at $N_c\ge3$. Since SU$(N_c)$ gauge group
is broken down to U(1)$^{N_c\!-1}$ by the vevs of adjoint scalars
there are $N_c\!-\! 1$ winding numbers, one per each U(1) factor.
Let us remind that the winding number
$n=0, \pm 1, \pm 2, \ldots$ counts the flux  of `magnetic' field in the ANO
 vortex
(it is an element of $\pi_1({\rm U}(1))\!=\!{\bf Z}$). Numerous winding
numbers lead to
existence of too many hadronic states in the spectrum~\cite{DS} (see also
\cite{HSZ} for the brane interpretation of this result). Namely, the
number of distinguishable families of $\bar{q}q$
meson states produced by \mbox{$p$-string/$(p-1)$-antistring} pairs
 (i.e. objects of
the type $[n_i]\!=\![0,\ldots,0,-1,1,0\ldots,0]$) is the integer part of
 $(N_c\!+\!1)/2$.

Another distinction, visible  already in the SU(2) theory~\cite{S},
is related with
higher winding numbers, $\!n\!>\!1$.
This  also produces an extra multiplicity in the hadron
spectrum if strings with higher winding numbers exist.
In QCD or in the large $\mu$ limit of the present theory we expect classification
of states under the center of the gauge group, ${\bf Z}_2$ for SU(2),
rather than ${\bf Z}$.

Consider as an example  the ANO vortex with two units of the flux, $n\!=\!2$. This 
string connects two quarks with two antiquarks producing an ``exotic''
state in the spectrum of the theory at small  $\mu$.
Note, that the string with $n\!=\!2$, in principle, can be
torn up by $W$ boson pair creation. It does not happen while energy is
less than
$2\,m_W\sim \Lambda$  which is a large quantity for
$\mu\ll\Lambda$ (here $\Lambda$ is the scale parameter of SU(2) theory).
We do not expect such exotic states to appear in QCD.

The discussion above of ``exotic'' states in the hadron spectrum
\cite{S} is based on the purely topological reasoning. Here
we consider the dynamical side of the problem. In fact, strings
with multiple $n$ are stable or unstable depending on the type
of the superconductor. Namely,  in the type I superconductor the
 tension $T_n$ of the vortex with winding number $n$
is less than $n\,T_1$ which is sum of tensions of $n$ separated  
vortices with the unit flux.
Therefore vortices  with $n\!>\!1$ are stable and we really have a
tower of ``exotic'' states in the spectrum.

On the contrary, in the type II superconductor vortices with $n_0\!>\!1$
are unstable against decay into $n$ vortices with the unit winding
number. Therefore, in this case ``exotic'' states are unstable  and
actually in the ``real world'' at strong coupling might not  be
observable at all.

The purpose of this paper is to find out the type of
superconductivity at the monopole point in the SU(2) Seiberg-Witten
theory perturbed by a small adjoint mass $\mu$. The result will allow us to make a
prediction about the presence of an ``exotic'' state in the hadron
spectrum of the theory associated with multiple winding numbers
of ANO vortices.

Note, that in this paper we consider the SU(2) gauge theory  and
discuss higher winding numbers $|n|\!>\!1$ in the single U(1) group. In
Refs.~\cite{DS,HSZ}  the SU$(N_c)$ gauge group is  considered and
$p$-strings with winding numbers $|n_p|\!=\!1$ in  $p$-th
U(1) factor are studied.  The string tensions are shown to
be proportional to $\sin(\pi p/N_c)$ which is interpreted as a
type I behavior of the $p$-string (see also the second  paper in Ref.~\cite{Sp2} 
for a
study of  a more general deformation of \ntwo theory).

It was already noticed \cite{HSZ} that the mass term for the
adjoint matter acts as a generalized Fayet-Iliopoulos \cite{FI}
term to  leading order in $\mu$ (see the next section for a
brief review). To this order, although $\mu\neq 0$, the extended \ntwo
supersymmetry is preserved in the effective low energy description
near the monopole point. Then  the superconductivity at the monopole
point looks as being   on the
border between  types I and  II and the ANO string looks like
BPS-saturated~\cite{HS}. However,  in~\cite{HSZ} it was conjectured
the BPS condition is spoiled by higher orders in the breaking parameter
$\mu/\Lambda$. In~\cite{Sp2} authors came to the same conclusion.

 In this paper we start by studying the effective theory in the vicinity 
of the monopole point. In leading order which accounts for linear in deviation 
terms we explicitly demonstrate that \ntwo supersymmetry is preserved. 
This preservation was firstly derived in Ref.~\cite{HSZ} within
the brane construction.
Our
consideration shows that together with the preservation of
\ntwo supersymmetry a certain U(1) flavor symmetry is also
preserved in the same order. In terms of the microscopic \none theory it is the $R$
symmetry broken only by quantum anomalies. In the effective theory an
anomalous nature of this U(1) shows up only in `nonperturbative' corrections of
order $\sqrt{\mu/\Lambda}$.

Next we study these corrections. Taken them into account  we show that
\ntwo SUSY is broken to \none and 
superconductivity at the monopole point turns out to be of 
type~I.
This result shows that the hadron spectrum of the theory looks
very different from what we expect in QCD. As we explained above
this means that  the tower of ``exotic''
states with multiple fluxes is present in the  hadron
spectrum.

The paper is organized as follows. In Sec.~2 we review the
confinement scenario near the monopole point. In Sec.~3 we
consider the leading order perturbation in $\mu$ and show
that \ntwo supersymmetry remains unbroken to this order. In
Sec.~5 and Sec.~6 we consider next-to-leading order corrections
in $\mu/\Lambda$. Sec.~7 contains our conclusions.

\section{Monopole condensation}

In this section we present a brief review on the monopole condensation
in the broken \ntwo gauge theory. 

In the \none superfield notations the Seiberg-Witten solution leads to the following
effective Lagrangian  for the dual U(1) gauge field $V_D(x,\theta,\bar\theta)$ and
 its \ntwo partner $A_D(x,\theta)$,
\begin{equation}
\label{swsol}
{\cal L}_{\rm eff}=
 \frac{1}{8\pi}{\rm Im}\!\int\!\!  {\rm d}^2\theta \,\tau_D(A_D)\,W^2_D-
\frac{1}{4\pi}{\rm Im}\!\int\!\!  {\rm d}^2\theta {\rm d}^2\bar\theta
\,\bar A_D\,A (A_D)\,.
\end{equation}
The solution is parametrized by vevs $a(u)=\langle A\rangle$, 
$a_D(u)=\langle A_D\rangle$ as functions of the modulus
$u=\langle\Tr\Phi^2\rangle$  where $\Phi$ is the the superfield describing the
adjoint matter in the original  microscopic SU(2) theory. 
In terms of these functions
\begin{equation}
\tau_D=-\frac{1}{\tau}=-\frac{\partial_u a}{\partial_u a_D}\,.
\end{equation}
Near the monopole point, $u=\Lambda^2$, the theory is in weak coupling regime,
${\rm Im}\,\tau_D \gg 1$, and 
the leading terms of of the expansions in $u-\Lambda^2$ 
are
\begin{eqnarray}
\label{sing}
&&
a_D=\frac{i}{2\Lambda}\left(u-\Lambda^2\right)\left(1-\frac{u-\Lambda^2}
{16\,\Lambda^2}\right)\left\{1+{\cal O}\left[\left(\frac{u-\Lambda^2}{\Lambda^2}\right)^2\right]
\right\}\,,\nonumber\\[2mm] 
&& a=\frac{4\Lambda}{\pi}+\frac{u-
\Lambda^2}{2\pi\Lambda}\left(1-\frac{u-\Lambda^2}
{16\,\Lambda^2}\right)\left(
\ln\frac{32\,\Lambda^2}{u-\Lambda^2}
+1+\frac{u-\Lambda^2}{32\,\Lambda^2}\right)\left\{1+
{\cal O}\left[\left(\frac{u-\Lambda^2}{\Lambda^2}\right)^2\right]
\right\}
\nonumber\\[2mm]
&& \tau_D=\frac{i}{\pi}\left(\ln\frac{32\,\Lambda^2}{u-\Lambda^2}+
\frac{u-\Lambda^2}{4\,\Lambda^2}\right)
\left\{1+{\cal O}\left[\left(\frac{u-\Lambda^2}{\Lambda^2}\right)^2\right]\right\}
\nonumber\\[2mm]
&&=\frac{i}{\pi}\left(\ln\frac{16i\,\Lambda}{a_D}-\frac{3i}{8}\,\frac{a_D}{\Lambda}\right)
\left\{1+{\cal O}\left(\frac{a_D^2}{\Lambda^2}\right)\right\}\,.
\end{eqnarray}
The logarithmic singularity at $u=\Lambda^2$ (i.e. at  $a_D=0$) is interpreted as 
the vacuum polarization of the monopoles whose mass $m_M=\sqrt{2}\,|a_D|$ 
vanishes at this point.

The logarithmic singularity is due to large distances of order $1/m_M\gg
1/\Lambda$, it makes  clear that the Lagrangian (\ref{swsol}) is {\em not}
Wilsonean. To make description local  one needs to ``integrate in'' light monopoles
what leads to  
\begin{eqnarray}
{\cal L}_{\rm Wils}\!\!\!&=&\!\! 
\frac{1}{8\pi}{\rm Im}\!\int\!\!  {\rm d}^2\theta \,\tilde\tau_D(A_D)\,W^2_D+
\frac{1}{4\pi}{\rm Im}\!\int\!\!  {\rm d}^2\theta {\rm d}^2\bar\theta
\,\tilde\tau_D(A_D)\,\bar A_DA_D
.\nonumber\\[2mm]
&&+\!\int\!\!{\rm d}^2\theta{\rm d}^2\bar\theta\left[\bar Q{\rm e}^{V_D}Q
+\bar{\tilde Q}{\rm e}^{-V_D}\tilde Q\right]
 + 2\,{\rm Re}\!\int\!\!{\rm d}^2\theta\,{\cal W}
\label{seff}
\end{eqnarray}
Here the \none chiral superfields  $Q$ and $\tilde Q$ describe the monopole
hypermultiplet. The third term in (\ref{seff}) is the kinetic term for these
fields, it includes also the gauge interaction of the monopoles. The superpotential
${\cal W}$ is then fixed by
\ntwo SUSY
\begin{equation}
{\cal W}=\sqrt{2}\;\tilde Q A_D\,Q\,.
\label{sup1}
\end{equation}
It reproduces correctly the monopole mass $m_M=\sqrt{2}\,|a_D|$. 

The Wilsonean Lagrangian (\ref{seff}) is also supplied with a normalization point 
$M_{\rm PV}$ which plays a role of the ultraviolet cut off for the $Q$, $\tilde Q$
loops.  This can be realized by introducing the Pauli-Villars regulators to the $Q$,
$\tilde Q$ fields  with the mass $M_{\rm PV}$. 
Then
\begin{equation}
\tilde\tau_D(a_D)=\tau_D(a_D)-\frac{i}{\pi}\ln\frac{i\,M_{\rm
    PV}}{a_D\sqrt{2}}
=\frac{i}{\pi}\left(\ln\frac{8\sqrt{2}\,\Lambda}
{M_{\rm PV}}-\frac{3i}{8}\,\frac{a_D}{\Lambda}\right)
\end{equation}
matches  the expression (\ref{sing}) after adding up the one-loop contribution of
$Q$, $\tilde Q$. The coupling $\tilde\tau_D$ is nonsingular at $a_D=0$, its 
value at
 $a_D=0$ 
\begin{equation}
\tilde\tau_D(a_D=0)=\frac{i}{\pi}\,\ln\frac{8\sqrt{2}\,\Lambda}
{M_{\rm PV}}\,.
\end{equation}
provides a weak coupling, ${\rm Im}\,\tilde\tau_D(0)\gg1$. 

Moreover,
for our further considerations  we can safely neglect by the
``nonperturbative'' $a_D/\Lambda$  term in
the expansion of 
$\tilde\tau_D$ limiting ourselves by 
$\tilde\tau_D(0)$. If the linear in $a_D/\Lambda$ correction were
 in the kinetic term of monopole fields it would be important. The
 absence of such corrections is seen from their absence in the
 logarithmic part of  $\tau(a_D)$ in Eq.~(\ref{sing}).

Now let us break \ntwo down to \none by adding mass term for the
adjoint matter in the microscopic SU(2) Seiberg-Witten theory
\beq
{\cal W}_{\rm mass}\ =\mu\,\Tr\Phi^2\,.
\eeq
We can treat this breaking perturbatively when $\mu\ll \Lambda$.
Then in the effective theory near the monopole point the perturbation 
$\Delta {\cal W}(A_D)$ of the superpotential
can be expanded in powers of $A_D$, 
\beq
\Delta {\cal W}(A_D)=\mu\, u\,(A_D)=\mu\,
\Lambda^2+\frac{i}{\sqrt2}\,\eta\, A_D+\frac{\mu_D}2\, A^2_D+{\cal
O}(\mu A^3_D/\Lambda)\ .
\label{sup2}
\eeq
Here $u(a_D)$ is the inverse to $a_D(u)$ function, and  the expansion
(\ref{sing})  for $a_D(u)$ leads to
\begin{equation}
\eta =-2\sqrt{2}\,\mu\,\Lambda\,,\qquad
\mu_D=-\frac{1}{4}\,\mu\ .
\label{etaxi}
\end{equation}
Note that $\mu$ and $\Lambda$ are complex quantities, their phases can be
changed by global U(1) rotations. In considerations below we will  often use this
freedom to fix parameters $\eta$ and $\mu_D$ to be real and positive.

Minimizing the superpotential given by the sum  of expressions (\ref{sup1})
and (\ref{sup2}) with respect to $Q$, $\tilde Q$ and $A_D$ we find that the Coulomb
branch shrinks to the point \cite{2}
$$
\langle a_D\rangle\ =\ 0\ ,
$$
while the monopole condensate appears
\beq
\langle \tilde q\,q\rangle\ =\ -\frac{i}{2}\,\eta\ .
\label{cond}
\eeq
Here $q,\tilde q$ are the scalar components of $Q,\tilde Q$.
The monopole condensation breaks the U(1) gauge group and leads to confinement
of electric charges. 

Parameters $\eta$ and $\mu$ in the superpotential (\ref{sup2}) play quite
different role in the effective QED description of the theory.
It was  noted in \cite{HSZ} that the linear term in (\ref{sup2})
is a generalization of the Fayet--Iliopoulos (FI) $D$-term.  Moreover, at
$\mu_D=0$ the $\eta$ term
 does not break \ntwo supersymmetry~\cite{HSZ}. We confirm this by explicit
calculations in the next section.

Breaking of \ntwo is due to the $\mu_D$ term in (\ref{sup2}). As we will show
 it is this term which shifts the mass of the field $A_D$ away from the photon
mass. Below considering  the effective QED on general ground  we view three
parameters $\eta$, $\mu_D$ and $\tilde\tau(0)$ as independent ones. Within the
Seiberg-Witten solution $\mu_D^2/\eta\sim\mu/\Lambda\ll 1$ in 
weak coupling.  

\section{Fayet-Iliopoulos term in \mathversion{bold} \ntwo QED}

In  this section  we consider the \ntwo QED theory given by Eq.~(\ref{seff}) and
perturbed by the superpotential~(\ref{sup2}) with $\mu_D=0$ and nonvanishing
$\eta$.  As it was mentioned above we neglect by ``nonperturbative'' corrections
$(a_D/\Lambda)^k$ and choose for simplicity $\tilde\tau_D(0)$ to be pure
imaginary  (i.e. taking the effective $\theta$ angle to be zero),
\begin{equation}
\tilde\tau_D(0)=\frac{4\pi\,i}{g^2}.
\end{equation}
The theory becomes the \ntwo QED where the $\eta$ perturbation appears as the
\ntwo  gene\-ralization of the Fayet-Iliopoulos term. We show that although
the $\eta$ perturbation does break the SU(2)$_R$ global symmetry associated with
\ntwo it does not break \ntwo  super\-symmetry.  A remarkable feature of this 
theory is that all particles enter one {\em long} supermultiplet of
\ntwo$\!$. It means that there are no nontrivial central charges in
this case in contrast with the Coulomb branch at $\eta=0$.

Having in mind vortex solutions we study then the reduction to 2+1 dimensions.
The reduced QED  has \nfour supersymmetry and we consider its superalgebra
with central charges. In the 2+1 dimensions strings are particle-like solutions
with nonvanishing central charges. Indeed, as we will see in
Sec.~4,  the ANO string is BPS saturated. It is in agreement with
Ref.~\cite{HSZ}.

\subsection{\mathversion{bold} \ntwo superalgebra in 3+1 dimensions}

Let us start rewriting the \ntwo QED in the component form. We will omit
below index $D$ referring to the dual variables such as $A_D$, $V_D$, $W_D$
and will use the terminology of the electric description. The Lagrangian in
component form then is
\begin{eqnarray}
\label{qed}
&&\! \! \! \! \! \! \! \! \! \! \! \!{\cal L}_{\rm QED}=-\frac1{4g^2}F^2_{\mu\nu}+
\frac1{g^2}|\partial_\mu a|^2
+ {\cal D}_\mu \bar \phi_f{\cal D}^\mu  \phi^f-U(\phi^f,a)  \\[1mm]
&&\! \! \! \! \! \! \! \! \! \! \! \!+\frac1{g^2}\lambda^f\, i\sigma^\mu
\partial_\mu\,
\bar
\lambda_f +
\psi\,  i\sigma^\mu {\cal D}_\mu \,\bar \psi +
\tilde\psi \, i\sigma^\mu {\cal D}_\mu\, \bar{\tilde\psi} 
 +\sqrt{2}\left[ i\,\psi \lambda^f \bar \phi_f
-\tilde\psi\lambda^f\phi_f -
\psi\tilde\psi\,a+{\rm h.c.}
\right]\nonumber
\label{qed31}
\end{eqnarray}
where  we introduce SU(2)$_R$ doublet $\phi^f$
for scalar  fields of the same electric (magnetic in the original dual
description above) charge
$+\,1$,
\begin{equation}
\phi^f=\left(\begin{array}{cc}
~q\\[1mm] \!\!-i\bar{\tilde q}
\end{array}\right)\;,
\end{equation}
(in our notation the bar is an equivalent of complex conjugation), and the
covariant derivative ${\cal D}_\mu = \partial_\mu-i\,n_e\,A_\mu$ ($n_e$
is the
electric charge). In the fermion sector
$\lambda_\alpha^f$ is the SU(2)$_R$ doublet of fermion fields in the gauge
supermultiplet (gluino and the fermionic partner of $a$),  while
$\psi_\alpha$ and
$\tilde \psi_\alpha$ are fermions from $Q$ and $\tilde Q$ superfields.
Note, that without $\Delta {\cal W}$ of Eq.~(\ref{sup2}) the full flavor
symmetry of the
\ntwo QED  is  SU(2)$_R\times$U(1)$_{\bar R}$. The U(1)$_{\bar R}$ charges of the
fields are given in the Table 1.
\begin{table}[h]
\begin{center}
\begin{tabular}{|c|c|c|c|c|c|c|}
\hline
~ & ~ & ~ & ~  &~ & ~ & ~\\[-0.1cm]
{\rm Fields} &  $a$ & $ \phi^f$  &  $A_\mu$ &  $\lambda^f$ &$\psi$ &$ \tilde\psi$
 \\[0.2cm]
\hline
\vspace*{-0.2cm}
~ & ~& ~  & ~  &~ & ~ & ~ \\
U$_{\bar R}(1)~{\rm charges} $ & -1 & $ 0 $ & $ 0 $& $-\frac{1}{2}$   &
$\frac{1}{2}$ &
$\frac{1}{2} $ \\[0.2cm]
\hline
\end{tabular}
\caption{ U(1)$_{\bar R}$ charges}
\label{tabU}
\end{center}
\end{table}

The scalar potential $U(\phi,a)$ in Eq.~(\ref{qed31}) has the form
\beq
U(\phi,a)=g^2 \,\Tr \left[\bar \phi_g \phi^f-\frac 1 2 \,\delta^f_g\, (\bar
\phi\phi)-\frac 1 2 \,\xi^f_g\right]^2+2|a|^2
\bar \phi_f\phi^f\ ,
\label{pinv}
\eeq
where $\xi^1_2=(\xi^2_1)^*=\eta$ while diagonal components of $\xi^f_g$ are
zero.  Adding up these components generalizes $\xi^f_g$ to the traceless
hermitian matrix  equivalent  to the triplet of SU$_R$(2),
\begin{equation}
\xi^f_g= \xi_a \left(\tau_a\right)^f_g\;, \qquad \xi_1={\rm Re}\,\eta
\;, \qquad \xi_2=-{\rm Im}\,\eta\;.
\label{xieta}
\end{equation}
The additional components $\xi_3=\xi^1_1=-\xi^2_2$ produce in the potential
$U$ the extra piece
\begin{equation}
U_{\rm FI}=g^2\left[-\xi_3\,(|q|^2-|\tilde q|^2)+\frac 1 2\,(\xi_3)^2\right]\;.
\end{equation}
It is easy to recognize $U_{\rm FI}$ as a component form of the
standard Fayet-Iliopoulos (FI) term~\cite{FI} in U(1) theory where in the
superfield notation it is $\xi_3\int {\rm d}^4 \theta \,V$.  We see that
the SU$_R$(2) symmetry unifies in the potential $U$ contributions from
$F$-terms
(due to linear in $A_D$ part of the superpotential $\Delta {\cal W}$)
with the FI
$D$-terms. For this reason we view pieces in $U$ associated with $\xi_a$ as
generalized  FI terms.

But this unification comes at a price:  the nonvanishing $\xi$ explicitly breaks the
SU(2)$_R$ invariance as we see from Eq.~(\ref{pinv}). 
 At first glance it suggests also breaking of the extended
\ntwo supersymmetry. However, as it was noted  in Ref.~\cite{HSZ} within the
brane construction, the extended
\ntwo supersymmetry is not broken by the FI terms. 
Within the effective field theory (\ref{qed}) we are able to verify this
explicitly by  invariance  under the following set of \ntwo
transformations:
\begin{eqnarray}
&&\delta A_{\alpha\dot\alpha}=-2i\,(\varepsilon^f_\alpha \,\bar \lambda_{\dot
\alpha f} +\bar\varepsilon_{\dot\alpha f}\,\lambda_\alpha^f)\;,\nonumber\\[1mm]
&&\delta
\lambda^f_\alpha=-\varepsilon^{\beta f}F_{\alpha\beta}+i\varepsilon^g_\alpha\,
D^f_g
+i\sqrt{2}\,\bar\varepsilon^{\dot\alpha f}\partial_{\alpha\dot\alpha}a\;,
\nonumber\\[1mm]
&&\delta
a=-\sqrt{2}\,\varepsilon^{\alpha f}\lambda_{f\alpha}\;,\nonumber\\[1mm]
&&\delta \phi^f=\sqrt{2}\,\varepsilon^{\alpha f}
\psi_\alpha+i\sqrt{2}\,\bar\varepsilon^{f}_{\dot\alpha}\bar{\tilde
{\psi}^{\dot\alpha}}\;,\nonumber\\[1mm]
&&\delta\psi_\alpha=-i\sqrt{2}\,\bar\varepsilon_{f}^{\dot\alpha}\,{\cal
D}_{\alpha\dot\alpha}\phi^f+2i \varepsilon^{f}_\alpha \phi_f\, \bar
a\;,\nonumber\\[1mm]
&&\delta\tilde\psi_{\alpha}=-\sqrt{2}\,\bar\varepsilon_{f}^{\dot\alpha}\,{\cal
D}_{\alpha\dot\alpha}\bar\phi^f-2 \varepsilon^{f}_\alpha \bar\phi_f\,
\bar a\;.
\label{trans}
\end{eqnarray}
Here $\varepsilon^f_\alpha,~\bar\varepsilon_{g}^{\dot\alpha}$ are eight 
parameters
of \ntwo transformations, and we use spinor notations for the Lorentz vectors,
$X_{\alpha\dot\alpha}=(\sigma^\mu)_{\alpha\dot\alpha}X_\mu$.
The generalized
$D$-terms $D^f_g$ are
\begin{equation}
D^f_g=-g^2\left[2\bar \phi_g \phi^f- \delta^f_g\, (\bar
\phi\phi)-\xi^f_g\right]\;.
\label{dterms}
\end{equation}
Technically, the preservation of \ntwo can be explained by noting
that the SUSY transformations of generalized $D$-terms (fixed by the
transformations of $\phi^f$, $\bar\phi_g$) do not depend on $\xi^f_g$
at all, i.e. the symmetry looks the same as at $\xi^f_g=0$.

Note that although the SU(2)$_R$ symmetry is broken by nonvanishing $\xi^f_g$
the U(1)$_{\bar R}$ is preserved at $\mu_D=0$.  As an explanation we suggest
that this U(1)$_{\bar R}$ can be related to the
$R$ symmetry of the microscopic theory which survives breaking by
$\mu\Tr\Phi^2$ and  broken only by quantum anomalies. As we will see in
Sec.~5 in the effective QED theory $\mu_D\neq 0$ breaks  U(1)$_{\bar R}$
reflecting the effect of anomalies.

The most general \ntwo superalgebra preserving the Lorentz invariance in
3+1 dimensions has the form:
\begin{equation}
\left\{Q^f_\alpha\,, \bar Q_{\dot \alpha
g}\right\}=2\,\delta^f_g \,\sigma^\mu_{\alpha\dot
\alpha}  P_\mu\;, \quad
 \left\{Q^f_\alpha\,,  Q_{\beta}^g\right\}=2\,\epsilon^{fg}\epsilon_{\alpha\beta}
{\cal Z}
\;,
\quad
 \left\{\bar Q^f_{\dot \alpha}\,,
\bar Q_{\dot\beta}^g\right\}=2\,\epsilon^{fg}\epsilon_{{\dot\alpha}{\dot\beta}}
\bar {\cal Z}
\;,
\label{alg}
\end{equation}
where $Q^f_\alpha\,, \bar Q_{\dot \alpha g}$ are supercharges ($\alpha,\dot
\alpha=1,2$ are Lorentz indices, $f,g=1,2$ are SU(2)$_R$ indices),
$P_\mu$ are
the energy-momentum operators, ${\cal Z}$ is the complex central charge.
It is well known that in gauge theories the real and imaginary parts of
${\cal Z}$
are related to electric and magnetic charges, and $|{\cal Z}|$ gives masses
of  charged BPS particles on the Coulomb branch of the Seiberg-Witten theory.
In the \ntwo QED it is particularly simple:
magnetic charges are absent and ${\cal Z}$ is given by $\int {\rm d}^3
S^\nu
\partial^\mu \,(a\, F_{\mu\nu})$ which reduces to the electric charge times
$\langle a
\rangle$. Here $S^\nu$ is the time-like three-dimensional hyperplane.

Few comments on representations of the algebra. As we mentioned above
the  central charge in Eq.~(\ref{alg}) gives masses of shortened
BPS hypermultiplets to be proportional to their electric charges. It
happens in the
Coulomb phase of the \ntwo theory at $\xi=0$. Once we switch on a
nonzero $\xi$ the charged matter fields $\phi^f$ develop vevs, see
Eq.~(\ref{cond}), and the theory is in the Higgs phase. Then electric
charges are
screened and the  central charge vanishes.  Therefore, there are no BPS particles
 at nonzero $\xi$, i.e. short supermultiplets are absent. Particularly,
photon enters
the long \ntwo multiplet containing eight bosonic states. Explicit
calculations of the
next subsection confirms this.

This consideration refers to particle-like objects in 3+1 and does {\em
not} include extended objects like
ANO strings. To consider string solutions which break the 3+1 Lorentz
invariance  it is
convenient to view them as  particle-like solitons in the world dimensionally
reduced  to 2+1. We will do this in Sec.~\ref{sec:2+1}.

\subsection{Mass spectrum}

In this subsection we explicitly work out the mass spectrum
 in the QED theory (\ref{qed}) to show that all particle states enter
one long  \ntwo
supermultiplet.  Consider scalar potential~(\ref{pinv}). By
SU(2)$_R$ rotation  we can always put it in the form where only
$\xi_3$ component is nonvanishing and positive,
\begin{equation}
U(q,a)\ =\ \frac{g^2}2\left(|q|^2-|\tilde q|^2-\xi\right)^2+
2g^2|\tilde qq|^2 +2|a|^2\left(|q|^2+|\tilde q|^2\right)\ ,
\label{uxi}
\end{equation}
where
\begin{equation}
\xi\ =\ \sqrt{\xi_a^2}>0\;.
\label{ksi}
\end{equation}
The $\xi$ dependent part of $U$ is nothing but mass terms for scalar
fields $q$ and
$\tilde q$ with $m_q^2=-g^2\xi$ and $m_{\tilde q}^2=g^2\xi$. At first
glance giving
masses to scalars but not to their fermionic partners does not preserve any
supersymmetry at all. However, the negative sign of $m_q^2$ signals
instability of $q=0$ point. Supersymmetry is restored  at the point of stability. 

Indeed, the potential (\ref{uxi}) is the sum of the positive-definite terms, and 
at the point where $U=0$ all $F$ and $D$ terms vanish. This happens at the
following vevs:
\beq
\langle q\rangle=\sqrt{\xi}\ ,
\quad \langle \tilde{q}\rangle =0\ , \quad  \langle a\rangle = 0\;.
\label{vevs}
\eeq
We choose   the field $q$ which develops vev to be real, it is just the unitary
gauge for the vector field.
Nonvanishing $\langle q\rangle$  breaks U(1) gauge group giving mass to the
photon (see Eq.~(\ref{qed})~),
\beq
m^2_\gamma\ =\ 2g^2\xi\ .
\label{photon}
\eeq
From the first term in the potential~(\ref{uxi}) we read off that the
mass of the real
scalar $q$ is the same as the photon mass,
\begin{equation}
m^2_q\ =\ m^2_\gamma
\end{equation}
Together the real $q$ field and three polarization of the massive photon
form up the boson part of the \none massive vector multiplet.  To extent that
\none is unbroken one scalar always has the same mass as the photon, no matter
whether the \ntwo SUSY is broken or not. Of course, there are also four
fermionic states in this multiplet.

What remains to calculate is masses of two complex field $\tilde q$ and
$a$ (four
bosonic states).  In the quadratic expansion of $U$ these fields  do not
mix with
others and their masses  are also equal to that of photon,
\begin{equation}
m^2_{\tilde q}\ =\ m^2_{a}\ =\ m^2_\gamma\;.
\end{equation}
Looking at the fermion part of the Lagrangian (\ref{qed31}) it is simple
to verify
that all fermion masses are also equal to $m_\gamma$.

We see that all particles (8 boson plus 8 fermion states) have the same
mass
$m_\gamma$. They form the long \ntwo multiplet. It is what we anticipate
based on on general consideration of \ntwo algebra.

\subsection{\mathversion{bold} \nfour superalgebra in 2+1 dimensions}
\label{sec:2+1}

Under dimensional reduction the \ntwo
superalgebra of eight supercharges $Q^f_\alpha$, $\bar Q_{\dot\alpha g}$
in 3+1 dimensions becomes  the
\nfour superalgebra in 2+1 dimensions. Our presentation here  follows
Ref.~\cite{St}.

The global symmetry group in 2+1 is SL(1,1;{\bf
R})$\times$SU(2)$_R$$\times$SU(2)$_{\bar R}$. The first factor here is the
Lorentz group in 2+1. Spinors of this group are real, so the difference
between
dotted and undotted indices disappears in 2+1. It results in the extra
global symmetry: the flavor U(1)$_{\bar R}$ we have
in 3+1D becomes SU(2)$_{\bar R}$ in 2+1D.

Fields of QED given by the dimensional reduction of the action
(\ref{qed})  to 2+1
form the following multiplets of the global symmetry
SU(2)$_R$$\times$SU(2)$_{\bar R}$. Bosonic fields of the \nfour gauge
supermultiplet are vector  potentials $A_\mu$ ($\mu$=0,1,2), which are
singlets, and the SU(2)$_{\bar R}$ triplet
 of scalars $a^{\bar f}_ {\bar g}$.  We use the matrix notation,
\begin{equation}
a^{\bar f}_ {\bar g}=\left(\begin{array}{cc} A_3&-\bar a \sqrt{2}\\ - a
\sqrt{2}&-A_3
\end{array}\right).
\end{equation}
for these scalars in the \{0,1\} representation.
The  third scalar appeared in the reduction  to 2+1 from the  spatial
component $A_3$ of the gauge field.

Fermion fields $\lambda^f_\alpha$,
$\bar\lambda^f_\alpha$ of
the gauge supermultiplet form the \{1/2,\,1/2\} representation of
SU(2)$_R$$\times$SU(2)$_{\bar R}$. We denote the quadruplet of these
fields as
\begin{equation}
\lambda_{\alpha \bar f}^f=\{\lambda_\alpha^f,\bar \lambda_\alpha^f\}\,, \qquad
f,\bar f=1,2\;.
\label{lff}
\end{equation}
This representation can also be viewed as a vector 
$\lambda^\mu_\alpha\,$,
$(\mu=1,2,3,4)$ of O(4) group by the relation
\begin{equation}
\lambda_{\alpha \bar f}^f= \lambda^\mu_\alpha \, (\tau^\mu)^f_{\bar f}\,
, \qquad
\tau^\mu=(i, \vec \tau)
\end{equation}
with  real fields $\lambda^\mu_\alpha$ (which are just real and
imaginary
parts of the original complex ones).
 In
the basis (\ref{lff}) the reality condition has the form
\begin{equation}
\left( \lambda_{\alpha \bar f}^f\right)^\dagger=\lambda_{\alpha f}^{\bar f}\equiv
\epsilon_{ff'}\epsilon^{\bar f \bar f'}\lambda_{\alpha {\bar f}'}^{f'}\;.
\label{reality}
\end{equation}

The matter hypermultiplet with the electric charge $+1$ contains scalars which
form the \{1/2,\,0\} representation -- the doublet  $\phi^f$, and
fermions in
\{0,\,1/2\} representation -- the SU(2)$_{\bar R}$ doublet $\psi^{\bar f}$,
\begin{equation}
\psi^{\bar f}=\left(\begin{array}{c}
\psi\!\\[1mm] \!\!\!-i\bar{\tilde\psi}\!
\end{array}\right).
\end{equation}
 The appearance of the
latter doublet is the only effect of the dimensional reduction for the
hypermultiplet.

In terms of fields introduced above the \nfour QED Lagrangian  in 2+1
takes the
form
\begin{eqnarray}
{\cal L}_{\rm QED}^{2+1}\!\!&=&\!\!\frac1{4g^2}\left[-F^2_{\mu\nu}+
\partial_\mu a^{\bar f}_{\bar g}\,\partial^\mu a^{\bar g}_{\bar f}
+2i\,\bar\lambda^f_{\bar f}\,\gamma^\mu \,\partial_\mu \lambda^{\bar
f}_f\,\right]
+ {\cal D}_\mu \bar \phi_f{\cal D}^\mu  \phi^f
+\bar \psi_{\bar f}\,\gamma^\mu \,{\cal D}_\mu \psi^{\bar f}
 \nonumber \\
&&-g^2\,\Tr \!\left(\bar \phi_g \phi^f -\frac 1 2\,\delta^f_g\,\bar \phi
\phi -\frac
1 2\,
\xi^f_g\right)^2 - \frac 1 2\, \bar \phi \phi \,\Tr \!\left(a^{\bar
f}_{\bar g}\right)^2
\nonumber \\[2mm]
&&-\bar \psi_{\bar f}\,\psi^{\bar g} \,a^{\bar f}_{\bar g}
+\sqrt{2}\left[ \psi^{\bar f} \gamma^0\lambda^f_{\bar f}\;\bar \phi_f+{\rm
h.c.}\right],
\label{qed4}
\end{eqnarray}
where $\bar \phi \phi=\bar \phi_f \phi^f$ and we use the Majorana basis
for the 2$\times$2
matrices $\gamma^\mu$:
\begin{equation}
\gamma^0=\sigma_2\,, \quad \gamma^1=i\sigma_3\,, \quad
\gamma^2=i\sigma_1\,.
\end{equation}
Notice that for fermionic fields $\bar \psi\equiv \psi^\dagger
\gamma^0$, i.e.
the bar denotes the Dirac conjugation. Notice also that to have
 the Majorana basis after dimensional reduction of the third
spatial direction we do the  spinor rotation
\begin{equation}
\psi^{\rm Majorana} =\frac{1+i\sigma_1}{\sqrt{2}}\, \psi\;.
\end{equation}
for all fermion fields, namely these rotated  fields are implied
starting from
Eq.~(\ref{qed4}).

 The supertransformations (\ref{trans}) after the reduction to 2+1 can be
 written as
\begin{eqnarray}
&&\delta A_\mu
\left(\gamma^\mu\right)_{\alpha\beta}=-\left(\varepsilon^{f}_{\bar
f}\right)_\alpha
\left(\bar\lambda_{f}^{\bar f}\right)_\beta
-\left(\bar\varepsilon^{f}_{\bar
f}\right)_\beta
\left(\lambda_{f}^{\bar f}\right)_\alpha
\;, \nonumber\\[1mm]
&& \delta a^{\bar f}_{\bar g}=-2\,\bar\varepsilon^{\bar
f}_f\,\lambda^f_{\bar g}+\delta^{\bar f}_{\bar g}\,\bar\varepsilon^{\bar
h}_f \,\lambda_{\bar h}^f
\;,\nonumber\\[1mm] &&\delta
\lambda^f_{\bar
f}=\frac i 2\,F_{\mu\nu}\,\gamma^\mu \gamma^\nu \varepsilon^{f}_{\bar
f}+ D^f_g\,\varepsilon^g_{\bar f}\,
-i\,\partial_{\mu} a^{\bar g}_{\bar f}\,\gamma^\mu\varepsilon^{f}_{\bar g}\;,
\nonumber\\[1mm]
&&\delta \phi^f=-\sqrt{2}\,\bar\varepsilon^{ f}_{ \bar f}\,
\psi^{\bar f}\;,\nonumber\\[1mm]
&&\delta\psi^{\bar
f}=-i\sqrt{2}\,{\cal D}_{\mu}\phi^f\,\gamma^\mu\varepsilon_{f}^{\bar f}\,
-2\sqrt{2}\,\phi^f\, a^{\bar f}_{\bar g}\,\varepsilon^{\bar g}_{ f}
\;,
\label{trans1}
\end{eqnarray}
where $(\varepsilon^{f}_{\bar
f})_\alpha=\{i\varepsilon^{f}_\alpha\,,i\bar\varepsilon^{f}_\alpha \}$
are eight
Grassmann parameters satisfying to the reality condition similar to
Eq.~(\ref{reality}).

The transformations (\ref{trans1}) imply that doublets of
 supercharges
$Q_\alpha^f$ and $\bar
Q_\alpha^f\equiv\epsilon^{ff'}(Q_\alpha^{f'})^\dagger $
are combined into the quadruplet
\begin{equation}
Q_{\alpha \bar f}^f=\{i\bar Q_\alpha^f,-i Q_\alpha^f\}\,, \qquad \bar f=1,2\;,
\label{qff}
\end{equation}
which forms one irreducible
representation \{1/2,\,1/2\} of SU(2)$_R$$\times$SU(2)$_{\bar R}$, the
same one
we have introduced above for $\lambda^f_{\alpha\bar f}$.

The transformations (\ref{trans1}) allow
to determine the \nfour superalgebra spanned by eight  supercharges
$Q^{f\bar f}_\alpha$.
The most general form of the \nfour superalgebra in 2+1 is
\begin{equation}
\left\{\left(Q^{f}_{\bar f}\right)_\alpha\,,  \left(\bar Q^{\bar
g}_g\right)_\beta\right\}=2\,\delta^f_ g \delta^{\bar f}_{\bar g}
\,\left(\gamma^\mu\right)_{\alpha\beta}
P_\mu-2\,\delta_{\alpha\beta}\left[
\delta^{\bar f}_{\bar g}{\cal Z}^f_{ g}+\delta^f_g \tilde{\cal Z}^{\bar
f}_{\bar g}
\right]
\;,
\label{alg4}
\end{equation}
where the traceless ${\cal Z}^f_{ g}$, $\tilde{\cal Z}^{\bar f}_{2 \bar g}$
are six central
charges. The  QED with generalized FI terms given by Eq.~(\ref{qed4})
leads to the
following expression for these central charges,
\begin{equation}
{\cal Z}^f_{g}= L\,\xi^{f}_{g}\int {\rm d}^2 S^\mu
F_{\mu}^*= 2\pi n \,L\,\xi^{f}_{g}
\;,\qquad \tilde{\cal Z}^{\bar f}_{\bar g}=0\;,
\label{cch}
\end{equation}
where $L$ is the total length of the reduced dimension, ${\rm d}^2 S^\mu$
is the element of the time-like two-dimensional hyperplane, and
$F_{\mu}^*=\epsilon_{\mu\nu\gamma}F^{\nu\gamma}/2$. The integral
$\int {\rm d}^2 S^\mu F_{\mu}^*$ is  the magnetic
flux which is quantized, $\int {\rm d}^2 S^\mu F_{\mu}^*=2\pi n$.

The \nfour
superalgebra (\ref{alg4}) with the central charges (\ref{cch})
generalizes  the
\ntwo algebra for the standard FI term~\cite{HS,GS}.  The generalization amounts
to introduction of SU(2)$_R$ group and  to substitution of the FI $D$-term
coefficient
$\xi_3$  by the matrix $\xi^f_g$. Nonvanishing $\xi^f_g$ explicitly breaks 
the SU(2)$_R$ flavor symmetry, the SU(2)$_{\bar R}$ is preserved, however,
because $\tilde{\cal Z}^{\bar f}_{\bar g}=0$.

To consider representation of the algebra (\ref{alg4}) let us write it down
in the rest frame where $P_\mu=(M,0,0)$ via four complex charges
\begin{equation}
Q^{f}_{\bar f}=\frac{1}{\sqrt{2}}\left[\left( Q^{f}_{\bar f}\right)_1 +
i \left(
Q^{f}_{\bar f}\right)_2
\right]
\end{equation}
instead of eight real $Q^{f\bar f}_\alpha$. For these supercharges with no
spinor index we
find from Eq.~(\ref{alg4})
\begin{equation}
\left\{Q^{f}_{\bar f}\,,  (Q^{\dagger})^{\bar g}_{g}\right\}=M\delta^f_g\delta^{\bar
f}_{\bar g} -\delta^{\bar f}_{\bar g} {\cal Z}^f_{ g} -\delta^f_g\tilde{\cal Z}^{\bar
f}_{\bar g}\;,\quad
\left\{Q^{f}_{\bar f}\,,   Q^{g}_{\bar g}\right\}=0\;,\quad
\left\{ (Q^{\dagger})^{\bar f}_f\,, (Q^{\dagger})^{\bar g}_{g}\right\}=0\;.
\label{rest}
\end{equation}
The positive definiteness of the first  anticommutator
implies the bound
\begin{equation}
M\ge{\rm Max}\,\left\{ \sqrt{{\cal Z}^{f }_{g}{\cal Z}_{f}^g/2}\,,\, 
\sqrt{\tilde{\cal Z}^{\bar f}_{\bar g}\tilde{\cal
Z}_{\bar f }^{\bar g}/2} \right\}
\label{bound}
\end{equation}
for the mass in 2+1D. If the central charges ${\cal Z}$ and $\tilde{\cal Z}$
are both nonvanishing the BPS multiplet saturating the bound consists  of
four bosonic and four fermionic states and preserves 1/4 of the 
\nfour SUSY. In our case when  \mbox{$\tilde{\cal Z}=0$} the bound $\sqrt{{\cal
Z}^{f }_{g}{\cal Z}_{f}^g/2}$  is saturated by BPS multiplet containing two bosonic 
plus two fermionic states,  1/2 of the
\nfour SUSY is preserved. 

The mass of the string in 3+1 dimensions is proportional to its length,
$M=TL$, and
Eq.~(\ref{bound}) becomes the bound  for the string tension $T$,
\begin{equation}
T\ge 2\pi |n|\,\xi\;.
\end{equation}
We used here expressions (\ref{cch}) for the central charges and the
definition
 $\xi=\sqrt{\xi^f_g \xi^g_f/2}$ introduced earlier in Eq.~(\ref{ksi}).
The BPS strings saturate this bound.

The BPS strings form the short (2 boson + 2 fermion) supermultiplet of \nfour.
From Eq.~(\ref{rest}) with
the central charges (\ref{cch}) one can see that
the invariant subalgebra is formed by four supercharges
\begin{equation}
\frac{1}{2}\left(\delta^f_g +\frac{1}{\xi}\,\xi^f_g\right) Q^{g}_{ \bar
f}\;, \qquad
\frac{1}{2}\left(\delta^g_f + \frac{1}{\xi}\,\xi^g_f\right)
(Q^{\dagger})_{g}^{\bar f}\;.
\label{supch}
\end{equation}
These supercharges annihilate the BPS states, 
we will use this for  construction of the BPS string solutions  in the next
section.

\section{The BPS solution for string}

In this section we consider the ANO string in QED with FI terms.  We  generalize
to the \ntwo supersymmetry (in 3+1D) the results of Ref.~\cite{DDT}
obtained in the
\none case  (see also \cite{GS}).

Let us start from reminding  basic facts about the ANO string in framework
of the {\em non-supersymmetric} Abelian Higgs model. The model contains
one complex scalar field $\varphi$ interacting with electromagnetic field,
\beq
S_{\rm AH}=\int {\rm d}^4x \left\{-\frac{1}{4g^2}F^2_{\mu\nu}
 +|{\cal D}_\mu \varphi|^2-\frac{\lambda}{2}\left(
|\varphi|^2-\xi\right)^2\right\}\,.
\label{ah}
\eeq
The field $\varphi$ develops vev, $\varphi^2=\xi$, the U(1) gauge group
is broken.
Photon acquires the mass, $m_\gamma^2=2g^2 \xi$, see Eq.~(\ref{photon}), the
Higgs mass is equal to  $m_H^2=2\lambda\xi$.

For an arbitrary $\lambda$ the Higgs mass
differs from that of the photon. The ratio of the photon mass to the
Higgs mass is an
important parameter, in the theory of superconductivity it characterizes
the type of
superconductor, see e.g. Ref.~\cite{dG}. Namely, for $m_H<m_\gamma$ we have the
type I superconductor, while for $m_H>m_\gamma$ we have the type II. This is
related to the fact that scalar field produces an attraction for two vortices,
while the electromagnetic field produces a repulsion.

The boundary separating superconductors of
the  I and II type corresponds to \mbox{$m_H\!=\! m_\gamma$}, i.e. to
the special
value of the quartic coupling $\lambda$,
\begin{equation}
\lambda\ = \ g^2\;.
\label{lambda}
\end{equation}
In this case vortices do not interact.
It is well known that vanishing of the interaction at $m_H=m_\gamma$ can be
explained  by the BPS nature of the ANO strings.  The ANO string
satisfy the first order equations and saturate the Bogomolny
bound~\cite{B}. This
bound follows from the following representation  for the string tension $T$,
\begin{equation}
T =2\pi\xi \,n+\int \!{\rm d}^2 x\left\{\frac{1}{2g^2}
\left[F_{12}+g^2
\left(|\varphi|^2-\xi\right)\right]^2
+ \left|\left({\cal D}_1+i {\cal D}_2\right)\varphi\right|^2 \right\}.
\label{tens}
\end{equation}
Here indices $1, 2$ denote coordinates transverse to the axis of the
vortex.
The minimal value of the tension is reached when  both  positive
terms in the integrand of Eq.~(\ref{tens}) vanish~\footnote{In the context of
the Landau-Ginzburg  approach to superconductivity the same system of first order
differential equations was derived by G.~Sarma at early sixties  in application 
to the case of
vanishing surface energy, see Ref.~\cite{dG}.},
\begin{equation}
F_{12}+g^2 \left(|\varphi|^2-\xi\right)=0\;,\qquad
\left({\cal D}_{1}+i{\cal D}_{2}\right)\varphi=0\;.
\label{ne1}
\end{equation}
 The string tension becomes
\beq
 T\ =\ 2\pi\xi \, n \ ,
\label{tension}
\eeq
where the winding number $n$ counts the quantized magnetic flux $\int \!{\rm
d}^2 x \,F_{12}=2\pi\,n$ (we assume that $n$ is positive).  The linear
dependence of
string tensions on $n$ implies absence of their interaction.

For the $n=1$ vortex it is convenient to
introduce the profile functions $s(r)$ and  $f(r)$,
\begin{equation}
\varphi(x) =\sqrt{\xi}\; s(r)\, {\rm e}^{i\,\alpha}\;,\qquad
A_n(x) =-\epsilon_{nm}\,\frac{x_m}{r^2}\, f(r)\ ,
\label{profil}
\end{equation}
where $r=\sqrt{x^2_1+x^2_2}$ is the distance and $\alpha$ is the polar
angle in the
(1,2) plane.  Then from Eqs.~(\ref{ne1}) we get
\begin{equation}
\frac1\rho\,\frac{{\rm d}f}{{\rm d}\rho} +s^2-1 = 0\ ,\qquad
\rho\, \frac{{\rm d}\,s}{{\rm d}\rho}+ (f-1)\,s= 0\ .
\label{foe}
\end{equation}
Here $\rho=g\sqrt{\xi}\,r$ is dimensionless distance.
The  profile functions are real
and   satisfy the boundary conditions
\begin{equation}
s(0)=f(0)=0\ ,
\qquad s(\infty)=f(\infty)=1\ ,
\label{bc}
\end{equation}
which ensures that the scalar field reaches its vev (\ref{vevs}) at the
infinity and the vortex carries one unit of the magnetic flux. The
equations (\ref{foe}) with the boundary conditions (\ref{bc}) lead to the
unique solution for the profile functions (although an analytic form of the solution is
not found).

Returning to the supersymmetric QED we see that the tension (\ref{tension})
coincides with the bound (\ref{bound}) derived from the \nfour algebra.
Moreover, vanishing of both terms in the integrand of Eq.~(\ref{tens}) is
in correspondence to annihilation of the bosonic solution by the action
of four
supercharges (\ref{supch}). Indeed, let us choose $\xi_{1,2}=0$,
$\xi_3=\xi$. Then,
the set (\ref{supch}) becomes $Q^{1\bar f}$, $\bar Q_{1\bar f}$. The
action of these
supercharges on bosonic fields can be read off supersymmetry transformations
(\ref{trans1}) for $\delta \lambda$ and $\delta \psi$ with following
substitution of parameters
\begin{equation}
\left(\varepsilon^{f}_{\bar f}\right)_\alpha
\to - \frac{i}{\sqrt{2}}\left[\delta^f_1\left(\delta^1_\alpha-i\delta^2_\alpha\right)
\bar\varepsilon_{\bar f }-
\delta^f_2\left(\delta^1_\alpha+i\delta^2_\alpha\right)
\varepsilon_{\bar f}\right]\;.
\end{equation}
Equating  $\delta \lambda$ and $\delta \psi$ for these transformations
to zero we
get
\begin{eqnarray}
&&F_{12} - D^1_1=0\;,\quad F_{01}=F_{02} =0\;,\quad D^1_2=D^2_1=0\;, \quad
\partial_{1}a^{\bar f}_{\bar g}=\partial_{2}a^{\bar f}_{\bar g}=0\;,
\nonumber\\[2mm]
&&\left({\cal D}_{1}+i{\cal D}_{2}\right)\phi^1=0\;, \quad
\left({\cal D}_{1}-i{\cal D}_{2}\right)\phi^2=0\;, \quad
\phi^f\, a^{\bar f}_{\bar g}=0\;.
\label{eqs}
\end{eqnarray}
Here $D^f_g$ is defined by Eq.~(\ref{dterms}) and we
imply that the bosonic configuration is time independent.
With our choice of
the $\xi$ orientation it follows from Eqs.~(\ref{eqs}) that
$\phi^2=a^{\bar f}_{\bar
g}=0$. Remaining equations coincide with Eq.~(\ref{ne1}) for Abelian Higgs strings
(up to correspondence $\phi^1 \to \varphi$).

Let us discuss now the fermion zero modes of the $n=1$ vortex.  We demonstrated
above
that the vortex configuration is annihilated by action of four (out of eight)
supercharges. Action of remaining four supercharges on the vortex  configuration
generates four zero modes. We get these modes substituting $\varepsilon^{f}_{\bar
f}$ as
\begin{equation}
\left(\varepsilon^{f}_{\bar f}\right)_\alpha
\to \frac{i}{\sqrt{2}}\left[\delta^f_1\left(\delta^1_\alpha+i\delta^2_\alpha\right)
\bar\varepsilon_{\bar f }-
\delta^f_2\left(\delta^1_\alpha-i\delta^2_\alpha\right)
\varepsilon_{\bar f}\right]\;.
\end{equation}
in the transformations (\ref{trans1}):
\begin{eqnarray}
\left(\lambda^{f}_{\bar f}\right)_\alpha\!\!\!&=&\!\! i\sqrt{2}\,
g^2\left(\xi-\varphi^2\right)\left\{
\delta^f_1\,
\varepsilon_{\bar f}\left(\begin{array}{c} \!\!1\!\\ \!\!i\! \end{array}\right)+
\delta^f_2\,
\bar\varepsilon_{\bar f}\left(\begin{array}{c} \!\!\!\!1\!\!\\ \!\! \!-i\!\!
\end{array}\right)\right\} \nonumber\\[1mm]
&=&\!\!  i\sqrt{2}\, g^2\xi\left(1-s^2\right)\left\{
\delta^f_1\,
\varepsilon_{\bar f}\left(\begin{array}{c} \!\!1\!\\ \!\!i\! \end{array}\right)+
\delta^f_2\,
\bar\varepsilon_{\bar f}\left(\begin{array}{c} \!\!\!\!1\!\!\\ \!\! \!-i\!\!
\end{array}\right)\right\} ,\nonumber\\[2mm]
\psi^{\bar f}_\alpha\!\!&=&\!\! i\,\bar\varepsilon^{\bar f}\left({\cal D}_1-i{\cal
D}_2\right)
\varphi\left(\begin{array}{cc} \!\!1\!\\ \!\!i\! \end{array}\right)=
i\,2\,\sqrt{\xi}\;\bar\varepsilon^{\bar f}\;\frac{(1-f)\,s}{r}\,\left(\begin{array}{cc}
\!\!1\!\\ \!\!i\! \end{array}\right).
\label{modes}
\end{eqnarray}

Four fermion zero modes obtained above are in correspondence with
shortening of the supermultiplet of vortices: two bosonic and two
fermionic states.
The shortening guarantees  the BPS nature of the
vortex and the exactness of the mass formula (\ref{tension}).

Our consideration above was done in the frame where the SU(2)$_R$ vector $\xi_a$
was oriented along the third axis, $\xi_{1,2}=0$. In this frame $\xi$
has the
meaning  of the standard FI $D$-term.  The case of FI $F$-terms when
$\xi_3=0$, $\xi_{1,2}\neq 0$ can be obtained from the case
of FI $D$-term by SU(2)$_R$ rotation.
In particular,  for the real parameter $\eta$ we have $\xi_1=\eta$ as the only
nonvanishing component (see Eq.~(\ref{xieta})). Then the above mentioned rotation
is the rotation around the second axis on the angle $\pi/2$.
For SU(2)$_R$ doublet $\phi^f$ this yields
\beq
\phi^f_{ F} \ =\ U^f_g \phi^g_{D}\ ,
\qquad
U\ =\ {\rm e}^{-i(\tau_2/2)(\pi/2)}=\ \frac1{\sqrt2}\left(\!\!\begin{array}{rr}
1&-1\\1&1\end{array}\!\!\right),
\eeq
where indices $D$ and $F$ denote fields in the theories with FI
$D$-term and $F$-term respectively. For the string solution it means
that
\begin{equation}
q_F=-i\bar{\tilde q}_F=\frac{\varphi}{\sqrt{2}}
\label{qF}
\end{equation}
with other fields unchanged.
The same SU(2)$_R$ rotation should be applied to fermion zero modes,
\begin{equation}
\lambda_{F \bar f}^f\ =\ U^f_g\,\lambda_{D \bar f}^f\;.
\label{Fmodes}
\end{equation}
 
\section{The mass term perturbation}

In this section in addition to the linear in$A_D$ term in $\Delta
{\cal W}$, see Eq.~(\ref{sup2}),
 we also switch on the mass term for $A_D$ (the term $\mu_D A_D^2/2$
in Eq.~(\ref{sup2})) in our low energy QED (\ref{seff}). It
 shifts the mass of the fermion field in the $A_D$ supermultiplet, 
and also
changes the scalar potential to the form
\begin{equation}
U(q,a)\ =\ \frac{g^2}2\left(|q|^2-|\tilde q|^2\right)^2
+ 2g^2\left|\tilde qq+\frac{i}{2}\,\eta+\frac{\mu_D}{\sqrt2}\,a\right|^2
+2|a|^2\left(|q|^2+|\tilde q|^2\right)\ .
\label{fift}
\end{equation}
In the covariant form (\ref{pinv}) an introduction of 
$\mu_D$  can be viewed as an addition  to the constant parameters $\xi^f_g$
terms $\Delta\xi^f_g$ which are linear in the fields $a$ and $\bar a$,
\begin{equation}
\Delta\xi^1_2=(\Delta\xi^2_1)^*=-i\sqrt{2}\,\mu_D \,a\,.
\label{dxi}
\end{equation}
The field $a$ has nonzero U(1)$_{\bar R}$ charge, see Table 1. In the 2+1D reduced 
theory $a$ is a component of the SU(2)$_{\bar R}$ triplet. Thus, adding $\Delta\xi$
of the form (\ref{dxi}) breaks the U(1)$_{\bar R}$ global symmetry (as well as
SU(2)$_{\bar R}$ in the 2+1D reduction).

Moreover, as we already mentioned, the mass term for $A_D$ breaks \ntwo
supersymmetry down to \none$\!\!$. We show this explicitly below
calculating the masses of boson fields. Later on we study what
happens to the ANO vortex at $\mu_D\neq 0$ and show that it becomes 
a non-BPS object.  

Note, that in the Seiberg-Witten theory 
the SU(2)$_R$ vector
$\Delta\xi$ contains part  along the vector $\xi$.  This is a generic situation.
In the special case of QED with the usual
$D$ type FI term an introduction of  $\mu_D$ produces $\Delta\xi$ which is
orthogonal to
$\xi$. In this case the vortex  remains BPS-saturated although \ntwo SUSY is
broken down to \none$\!$. The dynamical side of this property of QED with 
$D$-term is that the scalar field $q$ which gets vev
is in fact the superpartner of the photon in the massive \none multiplet,
so their masses are equal.

\subsection{Mass spectrum}

The minimum of the potential (\ref{fift}) coincides with the one in
obtained for $\mu_D=0$, i.e.
\begin{equation}
\left\langle q\right\rangle=i\left\langle \tilde
q\right\rangle=\sqrt{\frac{\eta}{2}}\;, \qquad
\left\langle a\right\rangle=0\;,
\label{fvac}
\end{equation}
where we for simplicity consider $\eta$ to be real and positive (it is always
possible to do by $U(1)$ rotation).
   Expanding the fields $q$,
$\tilde q$ and $a$ around their vevs  in the potential
(\ref{fift}) and extracting quadratic in fluctuations terms we get
after some algebra the $6\times6$ mass matrix
\beq
{\cal M}^2=\ m^2_\gamma\left(\begin{array}{cccccc}
1 & 0 & 0 & 0& \omega/{\sqrt2} & 0\\[1mm]
0 & 1 & 0 & 0 & \omega/{\sqrt2} & 0\\[1mm]
0 & 0 & 1/2 & 1/2  & 0 & \omega/{\sqrt2}\\[1mm]
0 &0 & 1/2 & 1/2 & 0 & \omega/{\sqrt2}\\[1mm]
\omega/{\sqrt2}& \omega/{\sqrt2}& 0 & 0 & 1+\omega^2 &
0\\[1mm]
0 & 0 &  \omega/{\sqrt2}& \omega/{\sqrt2}& 0 &
1+\omega^2 \end{array} \right),
\label{massmat}
\eeq
where $m^2_\gamma=2g^2\eta$ and we introduce the dimensionless parameter
\beq
\omega\ =\ \frac{g\,\mu_D}{\sqrt{2\,\eta}}\ ,
\label{omega}
\eeq
implying that $\eta$ and $\mu_D$ are real and positive.

Calculating the determinant of the matrix ${\cal M}^2-\lambda m^2_\gamma I$
we find the eigenvalues of the mass matrix. We have \beq
\det({\cal M}^2-\lambda m^2_\gamma I)=
-m^{12}_\gamma\lambda(1-\lambda)
\left[(1-\lambda)^2-\lambda\omega^2\right]^2\;.
\eeq
In the limit of small \ntwo breaking, $\omega\ll1$,  we have for six eigenvalues
\begin{equation}
m_i^2=m^2_\gamma\,\left(0,\, 1,\,
1-\omega,\,1-\omega,\,1+\omega,\,1+\omega\right)
\label{omsmall}
\end{equation}
We interpret this result  as follows. The massless state
is ``eaten up"  by the Higgs mechanism and the scalar with the mass
$m_\gamma$ is the scalar superpartner of photon in the massive
vector \none supermultiplet (see Sec.~3). A new phenomenon
occurs with other four real scalars. The \ntwo hypermultiplet is
split into two \none chiral fields (each contains two real
scalars) with different masses $m^2_\gamma\,(1\pm\omega)$. Thus, we
see explicitly that \ntwo supersymmetry is broken.

Consider also the opposite limit of strong \ntwo breaking,
$\omega\gg1$. We will use this limit in the next section. In this limit the
masses of a former \ntwo hypermultiplet are split as follows
\begin{equation}
 m_3=m_4=m_\gamma\, \omega=g^2\mu_{D}\ , \qquad
m_5=m_6=\frac{m_\gamma}{\omega}=2\,
\frac{\eta}{\mu_{D}}\ .
  \label{omlarge}
 \end{equation}
The limit $\omega\gg1$ can be
understood as taking the FI parameter $\eta$ to zero. Then it is
clear that the first set of masses, $m_{3,4}$, in
Eq.~(\ref{omlarge})  refers to the \none chiral field  $A$.
Its mass approaches the value $g^2\mu_{D}$ (note the $1/g^2$
normalization of the kinetic energy for $A$). Moreover, the
U(1) gauge symmetry is restored in the limit $\eta=0$ and the
four real scalars $(q,\tilde q)$ become massless, $m_1=0$,
$m_2\to0$, $m_5=m_6\to0$.

\subsection{The string: qualitative remarks.}

In this subsection we make a few qualitative remarks on the
ANO vortex solution for the theory with non-zero $\mu_D$.
First, let us note that the ansatz (\ref{qF}) with the field  $a=0$ does not
work any
longer. To see this consider the classical equation of motion for
$a$. We have
\begin{equation}
-(\partial_x^2+\partial_y^2) a +\sqrt{2}\,g^{4}\bar\mu_D\,\left(\tilde{q}q+\frac i
2\,\eta +\frac{\mu_{D}}{\sqrt{2}}a\right)+ 2g^2 a\left(\bar{q}q+
\bar{\tilde{q}}\tilde{q}\right)=0
\label{aeq}
\end{equation}
We are looking for the vortex solution for which  monopole fields
$q,\; \tilde{q}$ are not identically equal to their vevs and
acquire a non-trivial coordinate dependence. Then we see that
  $a =0$ does not satisfy  (\ref{aeq})   at non-zero $\mu_D$.
It is clear that now all three scalar fields $q,\; \tilde{q}$ and
 $a$ have a non-trivial  coordinate dependence.

We see that now the problem does not reduce to the simple
Abelian Higgs model of the type (\ref{ah}) with a single complex
scalar for which vortex solutions are well understood.
In principle, it is possible to develop a perturbation theory
around the $\mu_D=0$ solution in powers of $\mu_D$.
This is done in the recent paper \cite{ch}. The result of
numerical solution of equations of motion for vortex
in  \cite{ch}
seems to be in qualitative agreement with our conclusion
presented  in the next section.

 In the next section we take the following route.
 We find the explicit vortex solution for
a particular range of parameters $\eta$ and $\mu_D$ using our
freedom to vary them independently within our effective
QED description. The string tension in this limit  shows the type I
superconductivity.

Now let us make a comment about fermion zero modes of the
vortex. Since the \ntwo SUSY is broken down to  \none
  we have only four SUSY generators now.
It is clear on the other hand that the number of fermion zero modes of
the vortex
cannot jump as we switch on parameter $\mu_D $. Fermion zero modes
are given by small deformation of the ones in Eqs.~(\ref{modes}),
(\ref{Fmodes}).  Thus we have four fermion zero modes what corresponds to two
fermion states  in the supermultiplet of vortex states. It means that this
supermultiplet is {\em not} a
short one for \none and the vortex is
{\em not} BPS-saturated at non-zero $\mu_D$.

\section{Large $\mu_D$ limit}

 In this
section we use the possibility to consider $\eta$ and $\mu_D$ as an
independent parameters ignoring the relations (\ref{etaxi}), of
the Seiberg-Witten theory. Namely, we consider the limit of
large $\mu_D$, $\mu_{D}^2 \gg \eta$.  Note, that we still keep
$\eta\ll\Lambda^2$ and $\mu_D\ll\Lambda$ in order to use the weak
coupling QED description.

If the field $A$ is heavy we can integrate it out. From (\ref{sup1}) and
(\ref{sup2}) we get the effective  superpotential depending on
monopole fields $Q,\tilde Q$ only (see \cite{KSS,GVY} where the
similar integration is done for $\eta=0$)
\beq 
{\cal W}_{\mu_D\to\infty}=-\frac{ i}{\mu_D}\left(\tilde
QQ+i\,\frac{\eta}{2}\right)^2\ .
\label{suplm}
\eeq

The scalar potential of the Abelian Higgs model which follows
from this superpotential reads
\begin{eqnarray}
 U(q,\tilde q)_{\mu_D\to\infty}\ =\ \frac{g^2}2\left(|q|^2-|\tilde q|^2\right)\
+ \frac4{\mu_{D}^2}
\left(|q|^2+|\tilde q|^2\right)\left
|\tilde qq+i\,\frac{\eta}{2}\right|^2\ .
\label{potlm}
\end{eqnarray}
The first term here comes from the $D$-components of gauge
multiplet (see (\ref{fift})), while the second one comes from the
superpotential (\ref{suplm}).

The potential (\ref{potlm}) has a minimum at $q=\tilde q=0$ with
unbroken $U(1)$ gauge group as well as the one written down in
(\ref{fvac}) for $q$-fields with the broken gauge group. Below we
concentrate on the latter one.

Calculating the $4\times4$ mass matrix near this vacuum we
obtain one zero eigenvalue (corresponding to the  state ``eaten up" by
the Higgs mechanism), another one equal to $m_\gamma$, Eq.~(\ref{photon}),
 (the scalar superpartner of the photon), while other two masses are 
\beq
m_H=\frac{m_\gamma}{\omega}\ ,
\label{hmass}
\eeq
where $\omega$ is introduced  in (\ref{omega}).
This is the mass of two scalars in 
\none chiral multiplet. Note, as an
additional check, that the mass of a chiral field in (\ref{hmass})
coincides with $m_{5,6}$ in Eq.~(\ref{omlarge}), as
it is expected.

In order to use the effective potential (\ref{potlm}) we should
consider $\mu_D$ large enough to ensure that the mass of the scalar $a$ is
much larger than both photon mass $m_\gamma$ and the mass of
chiral field (\ref{hmass}). The mass of $a$ can be rewritten in
terms of $\omega$ as
\beq
m_a=g^2\mu_{D}\ =\ m_\gamma\omega\ .
\eeq We
see that (\ref{potlm}) is valid if
\beq
\omega\ \gg\ 1\ .
\eeq
We
impose this condition below in this section.

Now consider ANO vortex in the QED with the potential (\ref{potlm}). By
symmetry between $q$ and $\tilde q$ it is clear that the
classical solution corresponds to the ansatz with a single
complex scalar $\varphi$
\begin{equation}
q = \frac1{\sqrt2}\ \varphi\ , \qquad
\tilde q =- \frac{i}{\sqrt2}\ \bar{\varphi}\ .
\label{fans}
\end{equation}
Substituting (\ref{fans}) into (\ref{potlm})
 we arrive at the following Abelian
Higgs model
\beq
S=\int\!\!{\rm d}^4x\left\{\frac{1}{4g^2}F^2_{\mu\nu}+|{\cal D}_\mu\varphi|^2
+\frac{1}{\mu_{D}^2}\,|\varphi|^2\left(|\varphi|^2-\eta\right)^2\right\}.
\label{lmah}
\eeq
The mass of the scalar $\varphi$ near the vacuum
$\langle\varphi\rangle=\sqrt{\eta}$
coincides with the $m_H$ in Eq.~(\ref{hmass}).
 At $\omega\gg1$ we have
\beq
m_H\ \ll\ m_\gamma\ .
\label{t1con}
\eeq
Although the Higgs potential in (\ref{lmah})
 is not a standard one (see
(\ref{ah})) the condition
(\ref{t1con}) suggests that the superconductivity
in the model (\ref{lmah}) is of type I. We now show that this is
indeed the case.

The vortex solutions in the Abelian Higgs model with a standard
Higgs potential (\ref{ah}) under condition (\ref{t1con})
 have been
studied in \cite{Y}. To the leading order in $\log (m_\gamma/m_H)$
the vortex solution has the following structure: the
electromagnetic field is confined in a core with the radius
\beq
R_g\ \sim\ \frac{1}{m_\gamma}\log\frac{m_\gamma}{m_H}\ ,
\eeq
while the scalar field $\varphi$ is close to zero inside the core. 
Outside the core the electromagnetic field is very
small. At intermediate distances
\beq
R_g\ \ll\ r\ \ll\ \frac{1}{m_H}
\label{lregion}
\eeq
($r$ is the distance from the center of vortex in (1,2) plane)
the scalar field satisfies the free equation of motion. Its
solution reads \cite{Y}
\beq
s(r)\ =\ 1-\frac{\log \,(r\,m_H)}{\log\,
(R_g\,m_H)}\ ,
\label{hsol}
\eeq
where we assume the standard ansatz (\ref{profil}) for $\varphi$
and $A_\mu$. At large distances $r\gg1/m_H$ $\varphi$ approaches
its vev as 
\begin{equation}
|\varphi |-\sqrt{\eta}\sim\sqrt{\eta}\,
[s(r)-1]\sim \exp\,(-m_Hr)\ .
\end{equation}

The main contribution to the string tension comes from the
logarithmically large region (\ref{lregion}), where scalar field
is given by (\ref{hsol}).
 The result for the string tension is \cite{Y}
\beq
T\ =\ \frac{2\pi\,\eta}{\log (m_\gamma/m_H)}\ .
\label{stens}
\eeq
It comes from the kinetic energy of the scalar field in Eq.~(\ref{ah})
(the ``surface" energy).

Now it is clear that the vortex solution in the model
(\ref{lmah}) to the leading order has the same structure once the
condition (\ref{t1con})
is imposed. To see this observe that the main
contribution to the string tension comes from the region
(\ref{lregion}) where the scalar field is given by the solution
(\ref{hsol}) of free equations of motion. The details of the
scalar potential are not essential in this region. They become
important in the region $r\sim1/m_H$, but the ``volume" energy
coming from this region is suppressed by an extra powers of $\log
(m_\gamma/m_H)$ as compared with the one in Eq.~(\ref{stens}).

We conclude that at large $\omega\gg1$ our effective QED behaves
as a type I superconductor and the string tension of ANO vortex
is given by
\beq
T\ =\ \frac{2\pi\,\eta}{\log\,\omega}\ .
\label{lmtens}
\eeq

Now let us discuss what does this conclusion means for the
Seiberg-Witten theory. We have to reduce $\mu_D$ going to the
region $\mu_{D}^2\ll\eta$ where the relations
(\ref{etaxi}) are fulfilled. It is
clear that while we continuously reduce $\mu_D$ the string tension
$T$ stays below the BPS  value $2\pi\eta$, ensuring the type I
superconductivity. The reason for this is that the number of
states in the string multiplet cannot jump. As we discuss above
for $\mu_D^2\gg\eta$ the string is not a BPS state and
belongs to the ``long" supermultiplet.
Suppose that as we reduce $\mu_D$ the string tension crosses the
bound $2\pi\eta$ at some finite $\mu_D$. This would mean that
the string becomes BPS-saturated and should belong to the
``short" multiplet. This hardly can happen. Thus we conclude that
\beq
T\ <\ 2\pi\,\eta
\label{uneqten}
\eeq
for any nonzero $\mu_D$. The result (\ref{uneqten}) ensures the
type I superconductivity (see \cite{BV} where it is shown that
the string tension increases monotonically with the increase of
$m_{H}/m_{\gamma}$).

At $\mu_D=0$ the extended \ntwo supersymmetry is restored and the string
becomes BPS saturated reaching the bound $T=2\pi\eta$, see
Sec.~4. Let us emphasize that $2\pi\eta$ is the lower bound only when
$\mu_D=0$. Note,
that the number of states in the string supermultiplet still does not jump even at
$\mu_D=0$. Just the ``long" multiplet of \none SUSY becomes the ``short" multiplet
of
\ntwo SUSY.

Another way to reach the same conclusion is as follows. The type I 
superconductivity implies that there exist a  scalar field with nonvanishing vev
whose mass is below that of photon,  
see Eqs.~(\ref{omsmall}) and (\ref{omlarge}).
The reason for this is that
the scalar field with the lowest mass determines the large distance tail of the
vortex and ensures the attraction between different vortices.
For large $\mu_D$  the field $\varphi$
which form the string solution is the eigenvector of the mass
matrix with the lowest mass (\ref{hmass}).

Let us consider now the region of small $\mu_D$ and show that the
scalar fields which form the string solution have a non-zero
projection
on the state with the lowest mass. To the leading order in $\omega$
 two normalized eigenvectors of the mass
matrix (\ref{massmat}) with the lowest eigenvalue $m_{\gamma}^2
(1-\omega)$ are
\begin{equation}
\frac{1}{2}\left(\begin{array}{c}1\\1\\0\\0\\-\sqrt{2}\\0\end{array}
\right), \quad
\frac{1}{2}\left(\begin{array}{c}0\\0\\1\\1\\0\\-\sqrt{2}\end{array}
\right).
\label{eigf}
\end{equation}
Let us call by Higgs field
the combination of scalars which acquire non-zero vev at infinity.
It is  proportional to the vector
\begin{equation}
\frac{1}{\sqrt{2}}
\left(\begin{array}{c}1\\1\\0\\0\\0\\0\\ \end{array}\right),
\label{hinf}
\end{equation}
 see  (\ref{fvac}).
On the string solution this field changes from zero at the
origin to its vev $\xi$ at infinity.
Now it is easy to see that this vector has a
non-zero projection on the first eigenvector in Eq.~(\ref{eigf}).

\section{Conclusions}

In this paper we studied ANO vortices near the monopole point in
the Seiberg-Witten theory perturbed by the mass of the adjoint
matter. This perturbation reduces to FI $F$-term as well as to the
mass term of the $A_D$ field within the QED effective description. We showed
explicitly that FI term does not break \ntwo supersymmetry
and the ANO string is BPS-saturated if the mass parameter $\mu_D=0$. Then we break 
\ntwo SUSY by non-zero $\mu_D$ and show that the
superconductivity at the monopole point is of the type I.

Note, that although we referred to the particular perturbation 
by the adjoint mass, our results are valid for a generic perturbation.
Indeed, for any perturbation one expand the effective superpotential 
near the monopole points in powers of $a_D$. The crucial role is played by 
a linear term (generalized FI term) which ensures the 
monopole condensate. It is clear that an appearance of linear term
is generic, to have it zero one needs a very special perturbation.
The string is BPS in the linear approximation. Then the next quadratic term 
breaks \ntwo, so the corresponding central charge seizes to exist. 

In the broken \ntwo QED, which corresponds to the type I superconductor, the
ANO string tension satisfy the upper bound (\ref{uneqten}). In
particularly, in the limit of large $\mu_D$ the tension  is 
$2\pi\eta/\log \omega$, see Eq.~(\ref{lmtens}).
This means that vortices are bound by attractive forces. In
particular, the infinite towers of ``exotic'' hadron states
corresponding to all integer winding numbers $n$ of strings
emerge in the hadron spectrum.

Of course, if we increase $\mu$ and approach $\mu\sim
\Lambda$ the unwanted strings with $n>1$ become broken by the
$W$-boson creation. Moreover, as it is shown in \cite{HSZ}
within the brane approach the unwanted $(N_c+1)/2$
multiplicity of the spectrum in $SU(N_c)$ theory also disappears
at $\mu\sim\Lambda$, and quark can be connected with
antiquark by only one string. This string is believed to be
responsible for the confinement in the SQCD in the limit of
large $\mu$ \cite{W}.

However, this brane picture is hard to implement in the field
theory. One  reason for this is that at large $\mu$ dual
QED enters the strong coupling regime. Another one is probably
even more fundamental.
 The point is that the role of matter fields in the effective
QED (\ref{seff}) is played by monopoles. As $\mu$ approaches
$\Lambda$ the inverse mass of the dual photon (\ref{photon})
approaches
the size of monopole (which is of order of inverse $W$ boson mass,
$1/m_W\sim 1/\Lambda$). Under these conditions we hardly can
consider monopoles as local degrees of freedom and the dual QED
effective description breaks down.  In particular, we do not have
a field-theoretical description of $n=1$ string in the region of
large $\mu\ge\Lambda$.

On the other hand,  at small $\mu$ we have weak coupling QED
description, strings and confinement. However, we have an
infinite tower of unwanted states in the hadron spectrum.

\vspace{-7mm}

\subsection*{Acknowledgments}

\vspace{-4mm}

Authors are grateful to Gregory Gabadadze, Amihay Hanany, Mikhail Shifman and 
Matthew Strassler for helpful discussions. 
We are thankful to the Institute for Theoretical Physics
at Santa Barbara, where this work was initiated, for the support during  the SUSY99
program provided by the NSF grant PHY 94-07194.
A.~Y. would like to thank the Theoretical Physics Institute
at the University of Minnesota for hospitality and support.
The work of A.~V. is supported in part by DOE under the grant 
DE-FG02-94ER40823, A.~Y.
is partially supported by the Russian Foundation for Basic
Researches under grant No.~99-02-16576 and by the US Civilian Research
and Development Foundation under grant No.~RP1-2108.

\end{document}